\documentclass[11pt,twoside]{article}
\usepackage{asp2004}
\usepackage{psfig}
\usepackage{epsf}
\usepackage{graphics}
\usepackage{lscape}
\def\edcomment#1{\iffalse\marginpar{\raggedright\sl#1\/}\else\relax\fi}
\reversemarginpar

\begin{document}

\title{Ten Years of Super Star Cluster Research}

\author{Robert W. O'Connell}

\affil{Astronomy Department, University of Virginia, P.O. Box 3818,
Charlottesville, VA  22903-0818, USA}

\begin{abstract}

\vspace{0.1cm}

This conference demonstrated the tremendous progress that has been
made in the last decade in our observational and theoretical
understanding of compact, young, massive star clusters.  One of the
Hubble Space Telescope's main achievements was to show that these are
young analogues of classical globular clusters and form in profusion
in the local universe.  They are now recognized as key players in the
problems of galaxy evolution and star formation.  A major result of
the conference is the commonality of the star and cluster formation
process across an astonishing range of environments and scales.  With
the demise of HST, the focus of the field will probably shift to the
critical early phases of cluster formation and evolution as revealed
with new infrared and radio/mm facilities.

\end{abstract}
\thispagestyle{plain}

\section{Introduction}

The title of this conference summary is deliberately chosen.  I know
many have misgivings about the adjective {\it super}, which
is often associated with things you don't really want.  But some of the
star clusters we have been discussing are truly extreme and constitute
the most concentrated episodes of recent star formation outside the
nuclei of galaxies, The {\it super} appellation is also quite
venerable, having been first applied to clusters in the starburst
galaxy M82 by van den Bergh in 1971.  As for the time scale, intense
interest in such clusters was precipitated by a specific discovery
with the Hubble Space Telescope---that of a swarm of massive young
clusters in the central Perseus Cluster galaxy NGC 1275 by Holtzman et
al.\ (1992) only a little more than 10 years ago.  

The phenomenon of super star clusters (SSC's) was not unanticipated
before HST.  Ground-based observations had revealed a number of barely
resolved candidates for young, luminous clusters, mostly in actively
star forming galaxies.  These included M82-F (O'Connell \& Mangano
1978), NGC 2070/R136 in 30 Doradus (Melnick 1985), NGC 1569-A and -B
(Arp \& Sandage 1985), and the nuclear clusters of NGC 1705 (Melnick,
Moles, \& Terlevich 1985) and NGC 1140 (Gallagher \& Hunter 1987).
Remarkably, the clusters ranged in absolute magnitude up to $M_V
\sim -15$, 1000 times brighter than typical classical globular
clusters. Age estimates ranged from 10 Myr up to about 1 Gyr.  Even
one merger-remnant system of young clusters (NGC 3597) was known (Lutz
1991). 

HST's essential contributions were to show, first, that systems of
hundreds to thousands of SSC's could be detected in merger remnants,
e.g.\ in the reigning archetype NGC 4038-9 (Whitmore \& Schweizer
1995); and, second, that their luminosities and sizes were $L_V \sim
10^{6-8}\, {\rm L_{V,\odot}}$ and $R_e \la 5$ pc.  Combined with their
HST-determined colors and ground-based spectroscopic mass
measures of $M \sim 10^{5-6}\, {\rm M_\odot}$ (e.g.\ Ho \&
Filippenko 1996), the properties of the SSC's were exactly what was
expected for young analogues of classical globular star clusters. 

Thus, HST had demonstrated that {\it globular cluster formation
continues to the present epoch.}  This opened an exciting and
unexpected new window on a fundamental astrophysical process which had
been thought to be confined to protogalaxy conditions and therefore
impossible to study in detail in the local universe.  Because they
were luminous and detectable in a wide variety of environments, super
star clusters became recognized as key players in two distinct arenas
of the ``origins'' problems:  galaxy evolution and star formation.  The
result is the wealth of observational and theoretical contributions at
this conference.  

\vspace{0.1cm}

Observational capabilities were not a specific concern of the
conference, but there was a pervasive sense that we were approaching an
important crossroads.  The last several months have sharpened that
realization, and so I first want to discuss the observational
livelihood of massive cluster studies.


\section{The Demise of HST and New Observational Horizons}

The Hubble Space Telescope not only provided the trigger for this
field, it is indispensable to it.  The majority of important studies
of star clusters in the last decade, in our own or other galaxies,
have involved use of HST.  Although this conference concentrated
on young clusters, several talks reminded us that HST also
revolutionized the study of classical ($\ga 5$ Gyr old) globular
clusters.

The combination of two characteristics of HST was critical to its
success in star cluster research:  first (of course) is its
unprecedented spatial resolution (FWHM $\sim 0.04\arcsec$).  But
equally important is its access to the UV/optical bands
(0.1--1.0$\mu$) that contain the most diagnostics of stellar
populations.  While the near-infrared (1--3$\, \mu$) is a valuable
complement in population studies of the nearby universe, it cannot
equal the richness and power of the UV/optical region. 

The fundamental difficulty we face is that {\it HST's performance
will not be matched or superseded in the foreseeable future} either in
space or on the ground.  

At our November 2003 meeting, the prospect that HST would be
decommissioned in 2010 was sobering enough to impel discussion of how
best to use the ``little remaining'' HST observing time.  But in
January 2004, NASA cancelled further servicing of HST.  Without
hardware renewal, the estimated time to a 50\% failure probability for
scientific operations on HST is only about 30 months.  A sobering prospect
has become dire, and it is reasonable to assume that the
current golden age of star cluster observational science will come to
a close by 2007.

\vspace{0.1cm}

\noindent There are, nonetheless, other new observational horizons opening
up:

\vspace{0.08cm}

\noindent {\sl Radio, Millimeter, Sub-mm}:   Given steadily improving
sensitivities, these bands offer high-resolution
probes of the interstellar medium in and near {\it younger} SSC's.
They are essential for studying cluster formation.  The VLA/EVLA/VLBA
are mainly important for ionized gas and nonthermal supernova remnants.
ALMA (ca.\ 2007) offers unprecedented capability to study molecular gas at
resolutions of 0.01--0.1$\arcsec$.

\vspace{0.08cm}

\noindent {\sl The Spitzer Space Telescope} is a cryogenic 85-cm
telescope (3-4 year lifetime) for imaging and spectroscopy in the
3--160$\, \mu$ bands.  With a resolution of only about 2.5$\arcsec$ at
its shortest wavelengths, Spitzer is not well matched to cluster
scales outside our Galaxy.  However, it will make unprecedented 
contributions to our understanding of the dust environments of young
cluster-forming systems, the characteristics of heavily obscured
clusters, and most importantly the star formation process in our
own Galaxy. 

\vspace{0.08cm}

\noindent {\sl Large Ground-Based Telescopes}: The essential
contribution to date of large (6-m$\,+$) ground-based telescopes has
been high S/N spectroscopy of SSC's.  There are ambitious plans for
telescopes in the 15--30-m class.  With sophisticated adaptive-optics
systems, large telescopes can produce near diffraction-limited
performance ($\la 0.05\arcsec$) over modest fields of view in the near
IR---critical for probes of young, dusty clusters.  Because seeing
effects become worse at short wavelengths, there is little prospect
that AO can match HST resolution below $1\mu$.  However, optical
spectroscopy under the best ``natural seeing'' of 0.3--0.5$\arcsec$
will be quite powerful for studies of cluster dynamics, determination
of integrated ages/abundances from spectral synthesis, wind physics,
and related problems. 

\vspace{0.08cm}

\noindent {\sl JWST} is a 6-m, segmented mirror, near-IR optimized
space telescope planned for launch in 2011.  It will achieve excellent
IR imaging with FWHM $\sim 0.06 \arcsec$ (2.5$\times$ better than
HST/NICMOS) and free of the very bright atmospheric night-sky IR
background.  JWST will enable identification of SSC's in local dusty
environments and in principle to high redshifts ($z \ga 2$), assuming
they are sufficiently isolated from other structures.  It will support
IR color-magnitude imaging studies of populations in nearer clusters.
Its multi-object NIRSPEC will be an exceptionally powerful device for
obtaining IR spectroscopy of cluster stars, cluster environments, or
multiple SSC's in a 3$\arcmin$ field of view.  But it is doubtful on
both technical and budgetary grounds that JWST will offer HST-quality
resolution or sensitivity, and the corresponding leverage on stellar
population parameters in the nearby universe, much below 1$\mu$.

\section{Analysis Techniques}

Our ability to analyze stellar populations in distant clusters from
their integrated light is critically dependent on the availability of
high fidelity synthetic models for composite spectra and colors.
Fortunately, these are of a rapidly growing sophistication, and over a
dozen different model sets are now accessible over the Internet.
Model colors must include the effects of emission lines, which are a
strong function of metallicity.  We heard about the usefulness of
certain types of stars as diagnostics in particular age ranges:
Wolf-Rayet stars ($\sim 4$ Myr) and asymptotic giant branch stars
($\sim 100$ Myr) in particular.  Mechanical energy input from
Wolf-Rayet winds, a main determinant of the energy balance of a
cluster's interstellar medium and a driver of early gas outflows, is
also being thoroughly assessed through improved modeling.  An
important category needing more work are the red supergiants ($\sim
10$--20 Myr).  These can dominate the near-IR light of clusters toward
the end of the critical high ionization stage of evolution and provide
strong CO absorption signatures there, but model sets disagree
significantly concerning the details.

We were shown some beautiful hydrodynamic and N-body numerical
simulations.  These are coming to grips with many aspects of the
complexities of star formation within clusters, cluster and starburst
winds, and dynamical evolution.  It is fair to point to the things
left out of such simulations, but it also must be said that the
progress in 10 years has been as remarkable as on the observational
side. 

A major goal of this meeting was to bring theoreticians and
observers together, and a basic key to future success is obviously
to be relentless in closing the theory/observation loop.

\section{Star and Cluster Formation}

Perhaps the major result of the conference is the {\it commonality} 
of the star and cluster formation process across an
astonishing sweep of environments and scales, from the disk
of the Milky Way to massive cluster dominant galaxies like NGC 1275.
Dozens of studies revealed a number of basic features in common:

\begin{itemize}

\vspace{-0.15cm}

\item A large fraction ($\sim 50$--100\%) of all stars form in clusters.

\item The mass function for more massive clusters, regardless of age,
      is a power law $d\,n(m)/d\,m \sim m^{-2}$ and is approximately
      the same everywhere. 

\item Many systems produce super star clusters---namely, young
      ($\la 1$ Gyr), compact clusters with sizes comparable to
      classical globular clusters and masses in the range
      $10^5$--$10^8\, {\rm M}_{\odot}$.  These number in the
      thousands in extreme cases. The great majority of such
      identifications have been made at optical wavelengths in
      relatively low-extinction environments. 

\item Radio observations have located ultra-compact H II regions
      in many environments, most without optical counterparts.  These
      have high densities ($n_e \sim 10^6$ cm$^{-3}$) and pressures
      ($P/k > 10^8$) and small radii ($\la 5$ pc).  They are
      identified with very young ($\la 1$ Myr) SSC's still in a cocoon
      phase and powered by dozens to thousands of O stars.

\item The total number of clusters, the number of massive clusters, and
      the maximum cluster luminosity all appear to scale with the
      total star formation rate, with only a few exceptions (e.g.\ NGC
      1569).  This argues for a near-universal cluster formation
      process, in which the population of SSC's is governed by
      statistical effects. 

\item SSC's are predominantly single generation with an internal age
      spread of $\la 5$ Myr.  Exception: the nuclear clusters of late-type
      disk galaxies, which  experience more continuous star formation. 

\item The spatial structure of SSC's is the same in all environments,
      and is well fit by the Elson, Fall, \& Freeman (1987) profile, which
      has a power-law envelope without evidence of tidal truncation. 
      The effective radii of SSC's are comparable to those of classical
      globulars ($\sim 5$ pc).  Exception: some systems, e.g.\ NGC 1023,
      have populations of ``faint fuzzy'' star clusters with $R_e \sim 10$--20 pc.
      These may be products of special conditions prevailing in the disks of
      S0 galaxies. 

\item The initial mass function for massive ($\ga 5\, {\rm M}_{\odot}$)
      stars is a power law with $d\,n(m)/d\,m \sim m^{-{\alpha}}$ and
      $\alpha \sim 2\, \pm\, 0.3\,\,$ in all environments.  This is comparable
      to the Salpeter value ($\alpha = 2.3$).

\item The mass function for low-mass stars in SSC's, as inferred from
      the handful of kinematic studies to date, is consistent with the
      Kroupa (2002) function for the solar neighborhood in most
      cases.  This features a flattened power-law ($\alpha \sim 0$--1
      for $m \la 0.5\, {\rm M}_{\odot}$).  Exception: evidence for a
      possible deficiency of low-mass stars has been found in M82 (2
      of 3 clusters), NGC 1705 (nuclear cluster), and NGC 4038-9 (one
      cluster). 

\item The mass function of the molecular clouds that are the raw material
      for cluster formation is a power law with an index
      ($\alpha \sim 2$) matching that of the massive clusters.  
      Clouds in the Milky Way are often irregular in shape, and most
      may not be gravitationally bound, unlike the idealizations
      employed in older generations of models of star formation. 

\end{itemize}

\vspace{0.1cm}

\noindent As for the star formation mechanisms that underlie all this
activity, it was generally agreed that there is one theme operating on
these many scales with perhaps one major variation.  Many different
aspects of these mechanisms, oriented toward massive cluster
formation, were discussed and are summarized next.  Numerical
simulations illustrated fascinating details of the process, but
computational constraints prevented calculations for SSC-sized
systems.  The largest simulation shown was for 1000 ${\rm M}_{\odot}$. 

\begin{itemize}

\vspace{-0.15cm}

\item Formation of SSC's requires a high pressure medium, with $P/k
      \sim 10^{8-9}$, over 10$^4$ times higher than the prevailing
      pressure in the undisturbed ISM of the Milky Way disk.  It may
      be that high pressures must pervade a large volume of $\ga\,$kpc
      size.  We did not discuss at length the galactic-scale gas flows
      necessary to generate such regions or the sources and transfer of
      the turbulent energy required for the next step.

\item In the models presented, star formation is driven by supersonic
      turbulence, which must converge on the high pressure zones.
      Turbulence promotes hierarchical fragmentation of the parent
      cloud into gravitationally unstable clumps.

\item Once started, the time scale for star formation is short, only $\sim
      0.1$--1 Myr.

\item Theory predicts an extraordinary formation environment inside a
      protocluster, characterized by strong interactions and
      competitive accretion amongst protostars.  Those that survive
      longest in the denser gas clouds without dynamical ejection
      become the most massive.  The IMF resulting from numerical
      simulations of these processes resembles the Kroupa function.  
      The first observational evidence for an accretion disk around a
      massive protostar was reported for an O star in M17.  

\item The models (and observations, e.g.\ of young LMC clusters) show
      that mass segregation occurs rapidly ($\la$ a few$\times 10^5$
      yr), with the massive stars sinking to the center of dense
      clumps.  These environments may favor stellar coagulation to
      form yet larger objects.  The implication is that kinematical
      observations of even young clusters are affected by
      ``primordial'' mass segregation, leading to possible
      underestimates of cluster masses.  

\item The star formation efficiency (fraction of original cloud mass
      converted to stars) is $\sim 10$--20\% in the simulations.  

\item There is strong local feedback to the gas from ionization,
      stellar winds, stellar jets, and supernovae, but this is not
      well characterized yet and was not included in most models
      discussed.  Since lower-mass stars have long pre-main sequence
      lifetimes of $\ga 20$ Myr, they are vulnerable to feedback
      predations from their neighbors as long as they remain in the
      dense parts of the cluster.  The process that terminates star
      formation in proto-SSC's is not clear.  However, star formation
      efficiencies $\ga$10\% provide resistance to feedback and will
      probably produce a bound cluster.  

\item It was argued that magnetic fields have only a small influence
      on star formation compared to supersonic turbulence.  This is
      a major departure from earlier generations of models.  

\item Massive stars and massive clusters are both natural products of
      the kinds of hierarchical models discussed.  Special conditions
      are not required to generate SSC's, and these appear naturally
      as the upper end of the mass function is stochastically
      populated in larger molecular cloud complexes (a ``size of
      sample effect''). 

\item The major exceptions to this ``uniformitarian'' picture are those
      cases (notably NGC 1569 and 1705) where the cluster mass function
      is discontinuous and the SSC's are much more massive than other
      clusters.  A special formation mode probably operates in such
      circumstances, but the conditions that sustain it are not yet
      clear.

\item Good progress is being made in analyzing Pop II and Pop III star
      and cluster formation in the early universe ($z \ga 5$).  These
      studies are informed by the work discussed above on nearby
      systems.  In some ways, the situation is simpler since there are
      fewer free parameters (e.g. fewer complications from pre-existing
      structures).  However, it is critical to follow radiative
      cooling from H$_2$ and the rapidly changing complement of metals
      in full detail. Large scale HST/ACS surveys of classical
      globular cluster systems (e.g.\ Virgo) are quickly increasing the
      empirical test bed for such models.  The multiple subsystems of
      globulars revealed by HST observations are key avenues for
      understanding elliptical galaxy assembly.

\end{itemize}

\section{Galaxy Evolution}

Perhaps the overriding influence of SSC's on our view of galaxy
evolution has been to help dispel the classical picture that galaxy
and globular cluster formation were confined to a unique epoch within
$\sim 1$ Gyr of the Big Bang.  We now recognize that these are
continuing, hierarchical processes, extended in time to the present
epoch and strongly dependent on environment. 

Massive, compact clusters are now employed as one of the best tracers
of the star formation history of other galaxies.  As luminous, coeval
systems, they are readily identified (with sufficient resolution) and
relatively easy to age-date from broad-band colors.  They are
especially useful for diagnosing intense episodes of star formation
induced by mergers and other dynamical interactions.  Cluster age
statistics can isolate starburst events in time (with a precision of
$\sim \pm\, 0.2$--0.3 in $\log t$) and trace the propagation of recent
star formation across the faces of galaxies.  More continuous,
long-term star formation in galaxies can also be followed with cluster
statistics, as long as the important destruction mechanisms and
selection effects associated with cluster aging are taken into account.

Compact cluster destruction processes are well understood over long
time\-scales ($\ga 100$ Myr).  The main mechanisms are evaporation
from two-body relaxation (for lower mass clusters) and gravitational
shocking by the bulge or disk of the host galaxy.  For more massive
clusters, the slope of the mass function is preserved.  The enormous
statistical samples available in the case of systems like NGC 4038-9,
however, point to other important mechanisms on short timescales.  The
data show that roughly 90\% of the youngest clusters must be destroyed
(or at least drop below the surface brightness detection threshold)
over $\la 50$ Myr.  The same kind of ``infant mortality'' prevails
among young clusters in the Milky Way.  Early gas loss through stellar
feedback, which can unbind a cluster if its star-forming efficiency was
too low, is a likely driver, though the situation is not well
understood. 

SSC's can play a major role in modifying the interstellar medium of
galaxies.  For up to $\sim 50$ Myr, the stellar winds, supernovae, and
ionization from an individual massive compact cluster can have drastic
effects on the local ISM (on display in nearby systems like 30 Dor).
The crossing time for a composite cluster wind can be short enough
($\sim 1000$ years) that X-ray observations, for instance, reflect
contemporaneous internal conditions.  Young cluster winds will often
be enshrouded by dust, but near-IR spectral features such as the
Brackett lines have permitted detection of winds and diagnosis of their
properties in the emission line clusters of NGC 4038-9.  

In extreme environments such as the starburst core of M82, where
cluster separations may be only 2--5$\times$ their diameters, SSC's
will act {\it collectively} to reshape the global ISM of a galaxy.
They can drive galactic-scale winds, entraining large amounts of
external ISM material.  Outflows have now been identified on small
(the Milky Way center, the NGC 4038-9 ELC's), medium (blue compact
galaxies and dwarf galaxies), and large (M82, ULIRG's) scales.  In the
case of M82, for instance, the cluster winds have self-collimated to
produce the famous minor-axis gas plume.  

One of the most interesting species of SSC's are the nuclear clusters
found in a large fraction of all late-type (Scd--Sm) disk galaxies.
The nearest is in M33.  These have sizes and structures typical of
SSC's (much smaller than E galaxy cores).  Unlike SSC's, they contain
multiple generations of stars, indicative of approximately continuous
gas transfer from the disk.  Their origin and evolution are not well
understood.  Nuclear clusters in dwarf ellipticals may become
``ultra-compact dwarf galaxies'' if stripped from their parents during
an encounter and could also be mistaken for normal globular clusters
(as possibly is the case for $\omega\,$ Cen). 

One potential SSC formation environment not discussed at the conference
is the high pressure cooling flow region often found at the center of
X-ray clusters of galaxies.  There is much evidence for star formation
there, some of it induced by radio lobe/hot gas interactions.  The
presence of blue SSC's up to 60 kpc from NGC 1275 suggests that it
might be worthwhile to re-examine cooling flow formation mechanisms.  

A final important aspect of SSC's in galaxies is the possibility that
they are conducive to the formation of very massive stars and/or
intermediate mass black holes.  Numerical models are beginning to
explore the conditions necessary to generate massive objects by
coagulation following mass segregation in protocluster cores.  It will
be interesting to see if SSC's are plausible breeders of gamma ray
burst progenitors or might generate the seeds for the supermassive
black holes now recognized to be intimately linked to the evolution of
galaxy bulges.

\section{Conclusion}

There has been extraordinary progress during the last ten years
in our observational and theoretical understanding of massive
star clusters.  This field will no doubt continue to prosper
even after its observational mainstay, the Hubble Space Telescope,
is retired.  However, the focus will probably shift to the earliest
phases of cluster formation and evolution as revealed through
infrared and radio/mm observations.  It may well merge with
the concerted attack on the fundamentals of star formation, which
has just begun and for which the prospects over the next ten
years are exceptionally bright.

\end{document}